\documentclass[runningheads]{llncs}
\usepackage{cite}
\usepackage{amsmath,amssymb,amsfonts}
\usepackage{algorithmic}
\usepackage{graphicx}
\usepackage{textcomp}
\usepackage{xcolor}
\usepackage{hyperref}
\begin{document}

\title{Hades Again and Again: A Study on Frustration Tolerance, Physiology and Player Experience}

\author{Maj Frost Jensen \and
Laurits Dixen \and
Paolo Burelli\orcidID{0000-0003-2804-9028}}

\institute{Center for Digital Play, IT University of Copenhagen, Copenhagen, Denmark
\email{mfje,ldix,pabu@itu.dk}}

% \author{\IEEEauthorblockN{1\textsuperscript{st} ****}
% \IEEEauthorblockA{\textit{****}\\
% **** \\
% ****}
% \and
% \IEEEauthorblockN{2\textsuperscript{st} ****}
% \IEEEauthorblockA{\textit{****}\\
% **** \\
% ****}
% \and
% \IEEEauthorblockN{3\textsuperscript{st} ****}
% \IEEEauthorblockA{\textit{****}\\
% **** \\
% ****}
% }

\maketitle

\begin{abstract}
Accurately quantifying player experience is challenging for many reasons: identifying a ground truth and building validated and reliable scales are both challenging tasks; on top of that, empirical results are often moderated by individual factors. In this article, we present a study on the rogue-like game Hades designed to investigate the impact of individual differences in the operationalisation of player experience by cross-referencing multiple modalities (i.e., questionnaires, gameplay, and heart rate) and identifying the interplay between their scales.
\end{abstract}


\section{Introduction}\label{sec:intro}
Player experience modelling plays a crucial role in game development and game research by providing insights into how players perceive, interpret, and engage with games. It focuses on understanding gameplay's psychological, emotional, and cognitive aspects, allowing game designers and researchers to create more immersive and enjoyable experiences for players~\cite{wiemeyer_player_2016}. 

However, accurately quantifying different facets of player experience remains a challenging problem to this day for many reasons.
Player experience is inherently subjective and is influenced by various contextual factors, such as individual preferences, cultural background, and previous gaming experiences~\cite{tondello_player_2019}. Furthermore, it comprises multiple dimensions, including emotions, cognitive engagement, social interaction, and immersion. These dimensions interact with each other and are also influenced by personal differences~\cite{yannakakis_player_2013}. 

Finally, selecting appropriate tools and methods to quantify the experience of the players is challenging. Traditional survey-based methods may not effectively capture the nuances of subjective experiences and are summative in nature. Researchers need to develop innovative techniques, such as physiological measures (e.g., heart rate variability), behavioural observation, or eye tracking, to complement self-report measures and provide a more holistic understanding of the player experience~\cite{karpouzis_psychophysiology_2016}.

Most dimensions of the player experience, such as engagement, enjoyment, or frustration, all share these common challenges. For instance, some people have a high frustration tolerance and are less likely to get as annoyed or frustrated at minor setbacks, whereas those with a low frustration tolerance easily grow agitated at the same inconveniences. These individual differences are likely to have an impact on the operationalisation of frustration.

Within player experience research, these individual and contextual differences are often overlooked, as they rely on information that is often intangible and external to the games; however, aspects such as frustration tolerance have been studied extensively in other fields~\cite{meindl_brief_2019,seymour_frustration_2019}.

In this paper, we present a study that aims toinvestigate the moderation effect of contextual and individual differences on the player experience. In the study, we analyse and evaluate the player experience of a sample of players playing Hades~\cite{supergiant_games_hades_2020}, a competitive rogue-like action game, and investigate the interplay between a number of self-reported scales, players' performance, their heart rate, and their tolerance to stress and frustration.

With this study, we intend to shed some light on the following research questions:
\begin{itemize}
    \item[Q1] Do individual differences in frustration tolerance have a measurable impact on the player experience?
    \item[Q2] Do individual differences in the psychophysiological responses of players have a measurable relationship with the player experience?
\end{itemize}
Although the study is not intended to give a comprehensive answer to either of the two questions given the limitations of its scope and design; we selected the game and designed the study to maximise the likelihood of finding some evidence of these relationships, if they exist.

\section{Related Work}
Frustration and how it relates to gaming has been studied in many different contexts,this includes studies into how near-misses, despite causing frustration, can increase the urge for players to continue playing Candy Crush~\cite{larche_candy_2017}, adaptive design in video games where frustration plays a role~\cite{yun_o_2009}, as well as dynamic difficulty adjustment~\cite{zohaib_dynamic_2018}.

One of the primary challenges with employing frustration or other aspects of the player experience for adaptation of dynamic difficulty adjustment is the operationalisation of the chosen construct. 
Several studies have attempted to quantify frustration; among these, probably the most common approach is through questionnaires, with instruments such as the Game Experience Questionnaire (GEQ)~\cite{ijsselsteijn_game_2013}, Game User Experience Satisfaction Scale (GUESS)~\cite{phan_development_2016} or ENJOY~\cite{davidson_multi-dimensional_2018}.

These questionnaires offer powerful and validated tools to estimate several aspects of player experience; however, they heavily rely on players' self-awareness and offer low granularity in terms of which aspect of the game experience and the player behaviour has elicited a specific player reaction.

To alleviate these limitations, several researchers have investigated how to operationalise player experience through other means. For example, Shaker et al.~\cite{shaker_fusing_2013} investigate the fusing of gameplay data and head movement, while Burelli et al.~\cite{burelli_non-invasive_2014} used full-body posture and movement. Other researchers have investigated the use of psychophysiological signals, such as heart rate~\cite{drachen_correlation_2010} and galvanic skin response~\cite{christopoulos_body_2019}, showing how they can be powerful game-independent markers of player experience~\cite{dmello_generic_2011}.

While some of the aforementioned studies attempt to infer individual differences from data collected within the analysed gaming experience, most of them do not take into account contextual individual differences in terms of personality and attitude, which has been identified as a potential factor by Canossa et al.~\cite{canossa_arrrgghh_2011} and Chang et al.~\cite{chang_influences_2015}.
In this paper, we attempt to further explore this dimension of the player experience, estimate the impact of these contextual differences, and evaluate whether they can be measured and detected.

\section{Methods and Materials}
To estimate the interplay between personal attitude, reported player experience, and player behaviour, we conducted an empirical study of player experience in the game Hades~\cite{supergiant_games_hades_2020}. During the experiment, we collected self-reported feedback on the player experience using GEQ. Only the 'Core' and the 'Post-game' modules were used as the modules 'Social Presence' and 'In-Game' were not relevant to the experimental setup, as the game was played alone and we did not ask participants to answer questions during the gameplay. Instead, we collected in-game player behaviour and the players' heart rate through the Polar H10 ECG sensor\footnote{https://www.polar.com/uk-en/sensors/h10-heart-rate-sensor}.
The last aspect captured is the players' tolerance to frustration, estimated using the Behavioral Inhibition/Behavioral Activation Scale (BIS/BAS)~\cite{carver_behavioral_1994} questionnaire, paired with the Frustrative Nonreward Responsiveness (FNR)~\cite{wright_reduced_2009} questionnaire.

The BIS/BAS scales are primarily used to assess dispositions to anxiety, depression and other mental disorders, but have also proven to be helpful in finding relations in personality traits~\cite{vecchione_bis_2021}.
The questionnaire is structured into 4 sub-scales; BIS, BAS Drive, BAS Fun Seeking, and BAS Reward Responsiveness. BIS explains the tendency of people to avoid negative outcomes; this means that high BIS scores in individuals are related to higher feelings of anxiety.
BAS and its subscales explain individuals' tendency to respond to rewards or to engage in goals where the possibility of reward is there. FNR measures the motivational response to the lack of reward, acting as an extended subscale to BIS/BAS. Both scales are Likert scales with respectively 4 or 5 levels. 

The study includes 20 participants with a mean age of 27, mostly consisting of students from the IT University of Copenhagen and their acquaintances.
Before analysis, each gameplay video has been coded to identify the game events, detailed in the following sections. 
The collected data, the video coding, and the scripts used to analyse the data can be downloaded at \url{https://github.com/itubrainlab/hades\_player\_experience/}.

\subsection{The Game}
\begin{figure}[t!]
    \centering
    \includegraphics[width=0.9\textwidth]{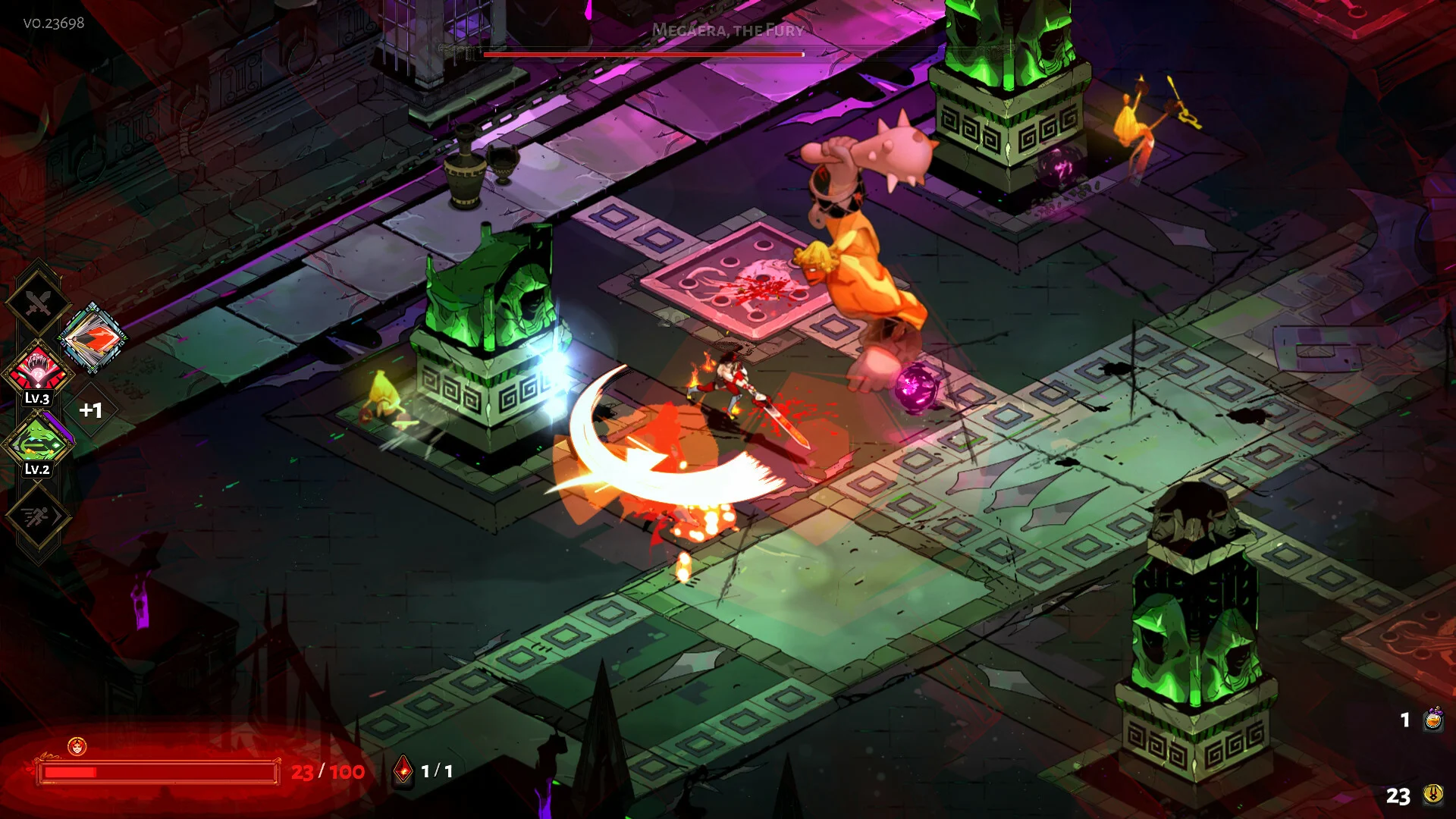}
    \includegraphics[width=0.9\textwidth]{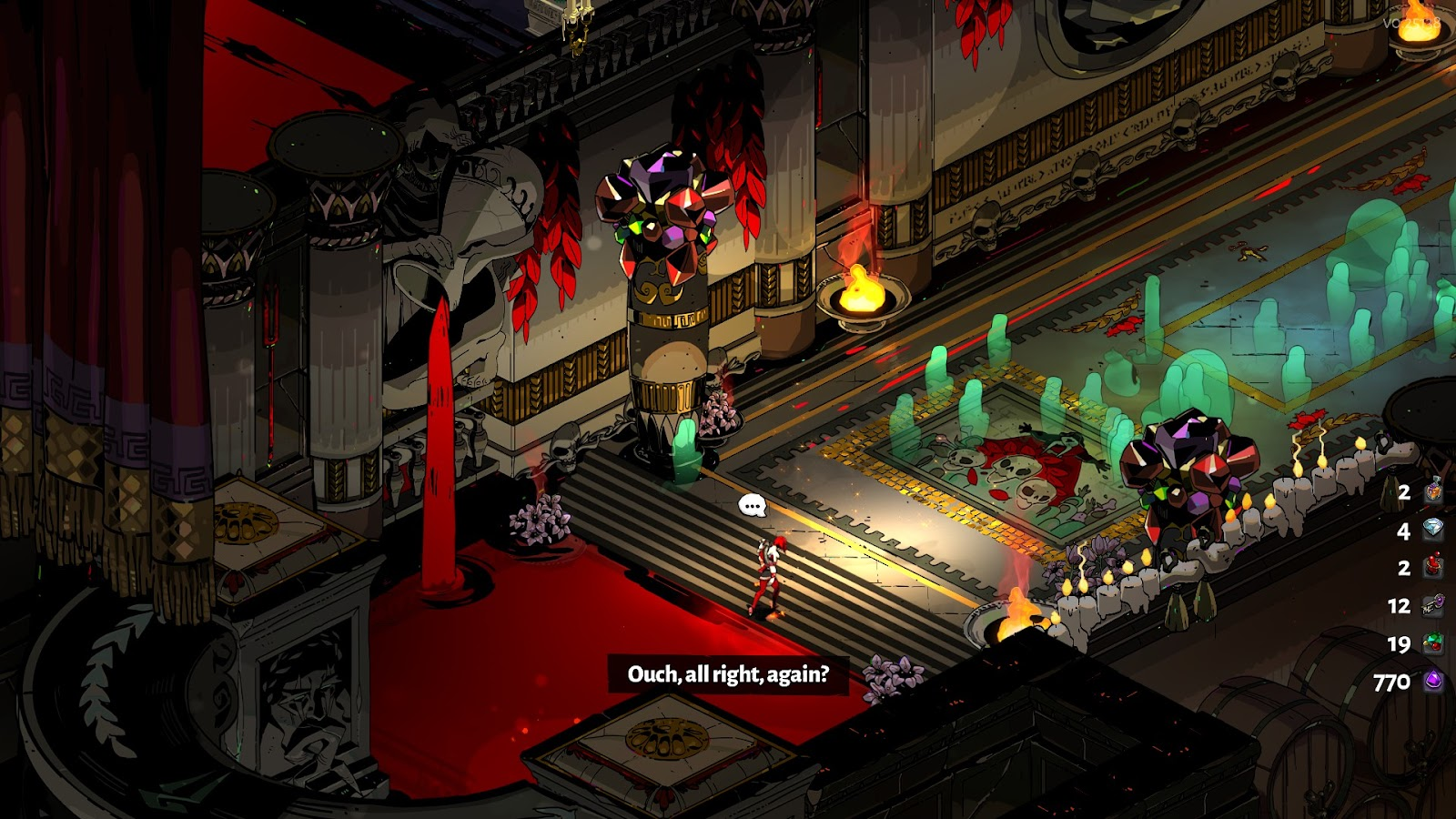}
    \caption{Screenshots of the Hades. Top: example of combat with multiple enemies, player is about to lose. Bottom: after losing the player character returns to the beginning again, losing almost all progress from the previous run. }
    \label{fig:game_pics}
\end{figure}

Hades is a single-player, rouge-like game in which the player proceeds through a series of progressively more challenging rooms, fighting enemies and collecting boons (Fig. \ref{fig:game_pics} Top). Death results in a total reset of the progress both in rooms and power-ups (boons) the player had accumulated during a run, and a return the the initial stage (Fig. \ref{fig:game_pics} Bottom).  In this way, the game is widely viewed as a challenging and potentially frustrating game to play, as any small mistake could cost the player a lot of progress.

We included only a single game, and any finding must be interpreted with this in mind. However, that also allows game-specific features to be extracted, which hopefully gives a more detailed understanding compared to general performance measures.

\subsection{Protocol}

Each recording session is individual and the player plays the first few levels of Hades for approximately 30 minutes using a joy-pad controller.  All sessions were conducted according to the following procedure:

\begin{enumerate}
        \item Setup the Polar H10 and make sure it connects with the Polar Flow app.
        \item Setup Hades and the controller.
        \item Test-run the recording software to make sure it works.
        \item Start a new save/profile on Hades on Hell mode.
        \item Invite the participant in and make sure they are comfortable.
        \item Ask them to put on the heart-rate monitor. Have a picture ready to show how it’s supposed to sit. Give them access to a bathroom and a towel so they can put it on in privacy.
        \item Have the participant sit down and make sure they are comfortable while connecting the sensor to the app.
        \item Let the player know that the first chamber will not contain enemies so they can run around and get used to the controller before continuing.
        \item Let the player know that there will be a menu at the start of each new run but all they need to do is click ‘Begin Escape’ when it pops up.
        \item Tell the player that once the 30 minutes are up, you will place a paper next to them stating "Last Run!" which means they continue to play until they complete the current, and then the Play Session will be over.
        \item Start the screen-recording software at the same time as the app.
        \item Sit for 1-3 minutes to capture the normal heart rate and then tell them they can begin playing.
        \item After the 30 minutes are up, discreetly place the piece of paper that says "Last Run!" next to the player, and wait for them to finish the current run.
        \item Let them know that they can remove the heart rate monitor and make sure that they are ready to complete the closing questionnaire.
        \item Bring up GEQ on the computer and let them fill it out.
\end{enumerate}

\section{Results}

\begin{figure*}[t!]
    \centering
    \includegraphics[width=\textwidth]{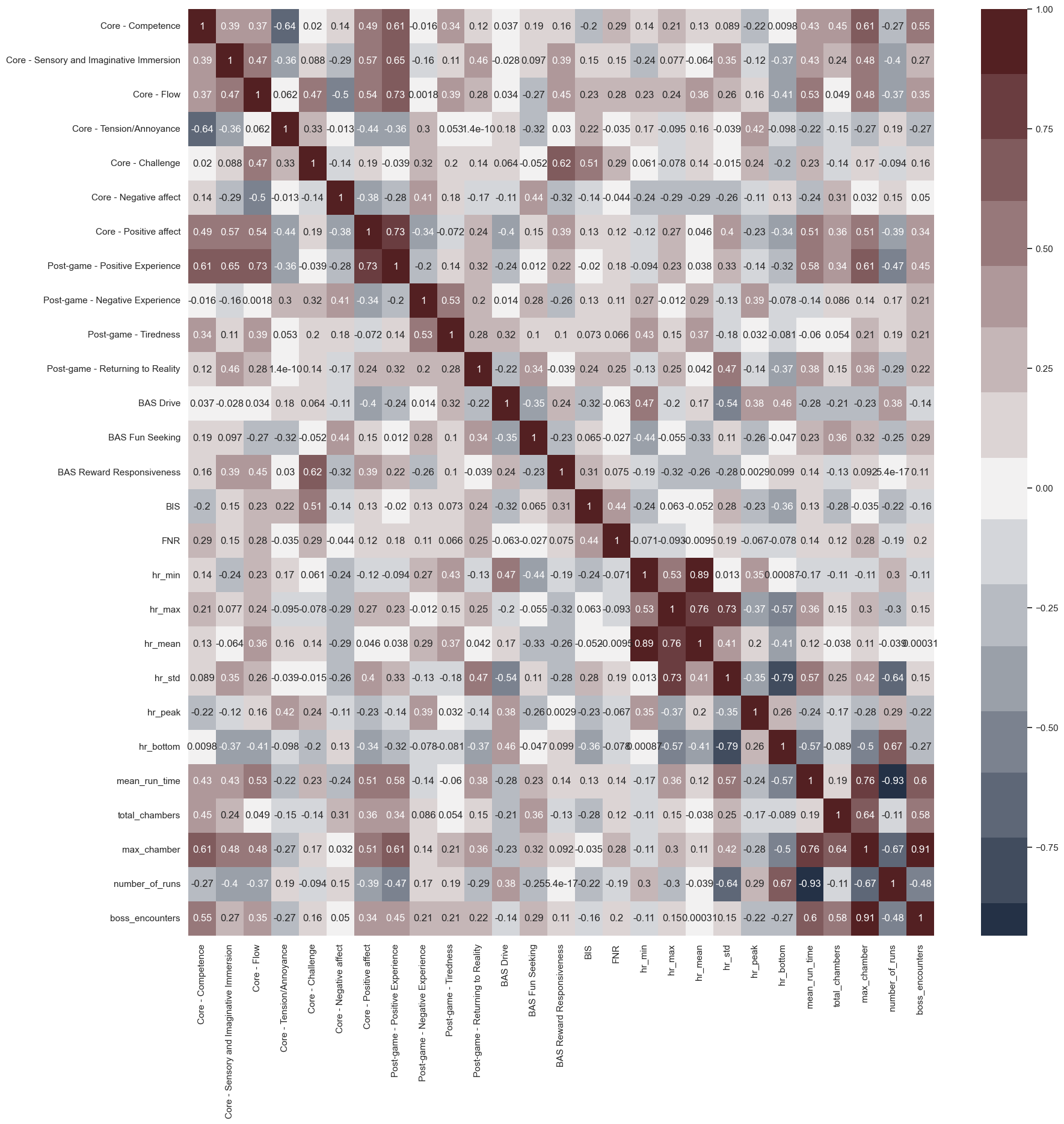}
    \caption{Correlation matrix (Pearson's) between the scales from the two questionnaires and the features extracted from the heart rate and game events.}
    \label{fig:corr_matrix}
\end{figure*}

\subsection{Features Extraction}

The heart rate data recorded is structured as a time series of heart rates recorded at a frequency of 1 Hz. From these data, for each player, we have extracted six characteristics capturing the minimum (\textit{hr\_min}), maximum (\textit{ hr\_max}) and mean (\textit{hr\_mean}) value of heart rate, its standard deviation (\textit{hr\_std}), and the fraction of time each player kept their heart rate around their minimum value (\textit{ hr\_bottom}) and around the maximum value (\textit{hr\_peak}).

The gameplay events extracted contain information about the duration of each run, the number of stages cleared, and the number of defeats in the game. The features extracted from the gameplay data are the time length of each run (\textit{mean\_run\_time}), the number of stages visited in total (\textit{total\_chambers}), the furthest stage reached in every run (\textit{max\_chamber}), the number of defeats and runs (\textit{number\_of\_runs}) and the number of times a player reaches the first boss encounter (\textit{boss\_encounters}).

\subsection{Analysis}

As a first approach to answer the research questions presented in Section~\ref{sec:intro}, we conducted a correlation analysis using Pearson's correlation ($r$) to identify the presence of a linear relationship between the scales measuring the frustration tolerance, the heart rate features, the in-game behaviour and the reported player experience.

Figure~\ref{fig:corr_matrix} shows the results of this analysis. The correlations show how each group of features and scales is internally correlated, with player experience aspects such as \textit{Flow} being strongly correlated with \textit{Positive experience} ($r>0.7$).

Across the scales, we can observe how the behavioural feature \textit{max\_chamber} shows a medium-strong correlation ($r>0.6$) with multiple scales in GEQ (\textit{Competence} and \textit{Positive Experience}) which is likely due to the tight relationship between enjoyment and performance in an action game like Hades.

It is interesting to observe how performance, described by \textit{max\_chamber}, demonstrates a medium correlation with both player experience scales and heart rate (\textit{hr\_std} and \textit{hr\_bottom}).
This relationship emerges also by dividing the players according to their performance. When the players are split according to the median value of \textit{max\_chamber} ($median=9$), the players with a higher performance than the threshold show a significantly higher \textit{hr\_std} ($7.4$ vs $4.6$) and a significantly lower \textit{hr\_bottom} (16\% vs 36\%) $p<.001$. 

Heart rate variability is an established measure of the activation of the sympathetic and parasympathetic nervous systems~\cite{sztajzel_heart_2004}, so a likely interpretation is that the capacity to increase and decrease the heart rate is a sign of the player's ability to respond to the challenge and perform better in the game.

The scores resulting from the frustration tolerance scales do not show any strong correlation with the other scale types and features; however, there is a medium correlation between \textit{Reward Responsiveness} and \textit{Reported Challenge} that could indicate the role of the individual importance of reward with their perception of challenge.

Another possible impact of frustration tolerance and heart rate on player experience is their potential role as moderators on other features, i.e. whether combining individual differences with gameplay features yields a stronger correlation with player experience.
To evaluate this kind of effect we trained a linear regression model using either heart rate or frustration tolerance features paired with gameplay features to assess the correlation of the linear combinations with reported player experience.

This analysis reveals that \textit{hr\_bottom} linearly combined with player performance (\textit{max\_chamber}) is strongly and significantly correlated with \textit{Competence} ($r>0.7$, $p<.001$) with a 16\% improvement over the single factor correlation shown in Figure~\ref{fig:corr_matrix}.
This result indicates that the differences in the dynamic behaviour of the heart rate play a moderator role in the players' feeling of competence.

In contrast, the same analysis performed on frustration tolerance scales does not reveal a strong combined correlation.

\section{Discussion and Conclusions}

In this paper, we present a study investigating the role of individual differences in the quantification of player experience. The individual differences accounted for in the study include players' heart rate behaviour and their self-reported response to frustration and stress, and the player experience is quantified through GEQ and the players' in-game behaviour.

The results and the analysis show no conclusive evidence of either an indirect or a direct impact of the reported frustration tolerance on the players' behaviour and player experience; however, the few reported medium correlations and moderator effects suggest this aspect might require some further investigation. It is important here to mention that the GEQ has received criticism that the factor structure is not stable ~\cite{law_systematic_2018}. In particular, the core factor \textit{negative affect} was problematic, as it was not clearly separated from similar factors \textit{challenge} and \textit{tension}. This is likely part of the explanation for why the results here are not clearer. We encourage further exploration of this topic in later studies. In particular, investigating how GEQ correlates with physiological measures and in-game behaviour in other settings than the one presented here would be an interesting undertaking.

Individual differences in heart rate behaviour, instead, show both a medium correlation with player performance and a strong correlation as a moderator effect between player performance and self-reported competence. 

These results are consistent with previous studies on the relationship between the activation of the autonomous nervous system and heart rate variability~\cite{sztajzel_heart_2004} and hint at the potential for heart rate dynamics to be a valuable feature for player segmentation and interpretation of players' experience.
However, heart rate standard deviation has limited granularity compared to heart rate variability (HRV)~\cite{sztajzel_heart_2004}, so to confirm the results of this study and draw a more accurate picture, we believe that further studies are needed using a measure with higher resolution. Additionally, this study was conducted on a particular game, Hades, to generalise findings; studies should include a broader range of games within the action game genre.

\bibliographystyle{plain}
\bibliography{references}

\end{document}